\documentclass[conference]{IEEEtran}
\usepackage{array}
\usepackage{multirow}% http://fctan.org/pkg/multirow

\usepackage[noadjust]{cite}
\usepackage{algorithm}
\usepackage{algorithmic}
\usepackage{amssymb}
\usepackage{bm}
\usepackage{amsmath}
\usepackage{cases} 
\usepackage{multirow}
\usepackage{booktabs}
\usepackage[caption=false,font=footnotesize]{subfig}
\usepackage{siunitx}
\usepackage{tabularx}
\usepackage{makecell}
\usepackage[table,xcdraw]{xcolor}
\usepackage{booktabs}

\ifCLASSINFOpdf
  \usepackage[pdftex]{graphicx}
  \graphicspath{{./}{./}}
  \DeclareGraphicsExtensions{.pdf,.jpeg,.png}
\else
  \usepackage[dvips]{graphicx}
  \graphicspath{{./}}
  \DeclareGraphicsExtensions{.eps}
\fi
\begin{document}

\title{An Efficient Partial Sums Generator for Constituent Code based Successive Cancellation Decoding of Polar Codes 
\vspace{-0.5em}}

\author{\IEEEauthorblockN{Tiben Che and Gwan Choi}
\IEEEauthorblockA{Department of Electrical and Computer Engineering\\
Texas A\&M University, College Station, Texas 77840\\
Email: $\lbrace$ctb47321, gwanchoi$\rbrace$@tamu.edu}}


%Telephone: (800) 555--1212, Fax: (888) 555--1212}


\maketitle

\begin{abstract}
%\boldmath
This paper proposes the architecture of partial sum generator for constituent codes based polar code decoder.
Constituent codes based polar code decoder has the advantage of low latency. 
%However, none of the specific designed partial sum generator for that has been investigated so far. 
However, no purposefully designed partial sum generator design exists that can yield desired timing for the decoder.  
We first derive the mathematical presentation with the partial sums set $\bm{\beta^c}$ which is corresponding to each constituent codes. 
%This results in a shift register based partial sum generator. 
From this, we concoct a shift-register based partial sum generator.
Next, the overall architecture and design details are described, and the overhead compared with conventional partial sum generator is evaluated. 
Finally, the implementation results with both ASIC and FPGA technology and relevant discussions are presented. 
\end{abstract}

% \fi
\IEEEpeerreviewmaketitle

\section{Introduction}
\label{Introduction}
Recently, polar code \cite{arikan2009channel} has received increasing attentions because it is the first code which provably achieves the channel capacity. 
Its low-complexity encoding and decoding schemes make it very promising for practical application. 
There are three widely known algorithms for polar codes decoding.
E. Arikan in~\cite{arikan2009channel} presents a successive cancellation (SC) algorithm which can successively accomplish decoding with recursive cancellation. 
I. Tal~\cite{tal2011list} makes the SC algorithm more competitive by exploring more paths among the codewords tree; this method is referred as list successive cancellation (LSC).    
Also, N. Hussami et al. in~\cite{hussami2009performance} shows that the belief propagation (BP) can be applied as decoding algorithm.

Although many efforts have been made for BP decoder~\cite{xu2015xj},~\cite{yuan2013architecture} and~\cite{lin2016high}, the BP decoding still suffers from the problem of high computing complexity. 
Thus, SC and LSC attract more studies especially on their hardware architecture~\cite{leroux2013semi}~\cite{zhang2013low}~\cite{yuan2014low}~\cite{che2015overlapped}~\cite{balatsoukas2014hardware}~\cite{balatsoukas2014llr}. 
SC decoding is based on the feedback, which is also called partial sum, from decoded codewords. 
A partial sum generator (PSG) is needed for each SC decoder. The partial sum needs to be calculated at the same clock cycle when the codewords are determined. 
Thus, the calculation of partial sum is on the critical path of the decoding and can affect the maximum frequency of the decoder. 
Some works have been done for a good PSG design.
C. Leroux~\cite{leroux2013semi} proposed an indicator function based PSG (IF-PSG). 
C. Zhang~\cite{zhang2013low} proposed a PSG with feedback part (FB-PSG).  
J. Lin~\cite{lin2015hybrid} proposed a hybrid PSG for LSC. 
G. Berhault proposed a shift-register-based PSG (SR-PSG)~\cite{berhault2013partial}~\cite{berhault2015partialsum}, which is able to increase the timing performance and reduce the hardware complexity. 
Y. Fan~\cite{fan2014efficient} proposed a similar architecture with SR-PSG however with higher level simplification. 

Both SC and LSC suffer from the long latency problem. The constituent code based decoding has been studied recently since it is capable of significantly reducing decoding latency~\cite{alamdar2011simplified}~\cite{sarkis2014fast}~\cite{che2016tc}.
All the aforementioned PSGs are capable of increasing the timing performance of SC decoder. 
However, none of them has considered the constituent codes based decoding. 
Since introducing the concept of constituent codes into decoding processing can significantly reduce the latency, it is reasonable and necessary to design a constituent-codes-compatible PSG. 
In this paper, we propose an efficient PSG for constituent code based SC decoding, and this is the first architecture of PSG for constituent code based SC decoder. 
First, we derive the mathematical presentation for constituent based PSG. This derivation is based on the SR-PSG for conventional SC decoder. 
Next, the overall hardware architecture and design details are proposed.
The timing and hardware complexity are evaluated as well. 
Finally, the implementation result are presented. This architecture is implemented with both VLSI and FPGA technology.
The relevant discussions are also mentioned as well.

This paper is organized as follows. 
The relative background is reviewed in section~\ref{Background}. 
In following, the proposed design including the mathematical derivation are described in section~\ref{Proposed Design}. 
After that, the implementation results and reverent discussions are presented in section~\ref{Implementation results and discussions}. 
Finally, this paper is concluded in section~\ref{Conclusion}.

\section{Background}
\label{Background}
\subsection{Polar Code}
\label{Polar Code}

\begin{figure}[!t]
\centering
\subfloat[]{\includegraphics[width=1.6in]{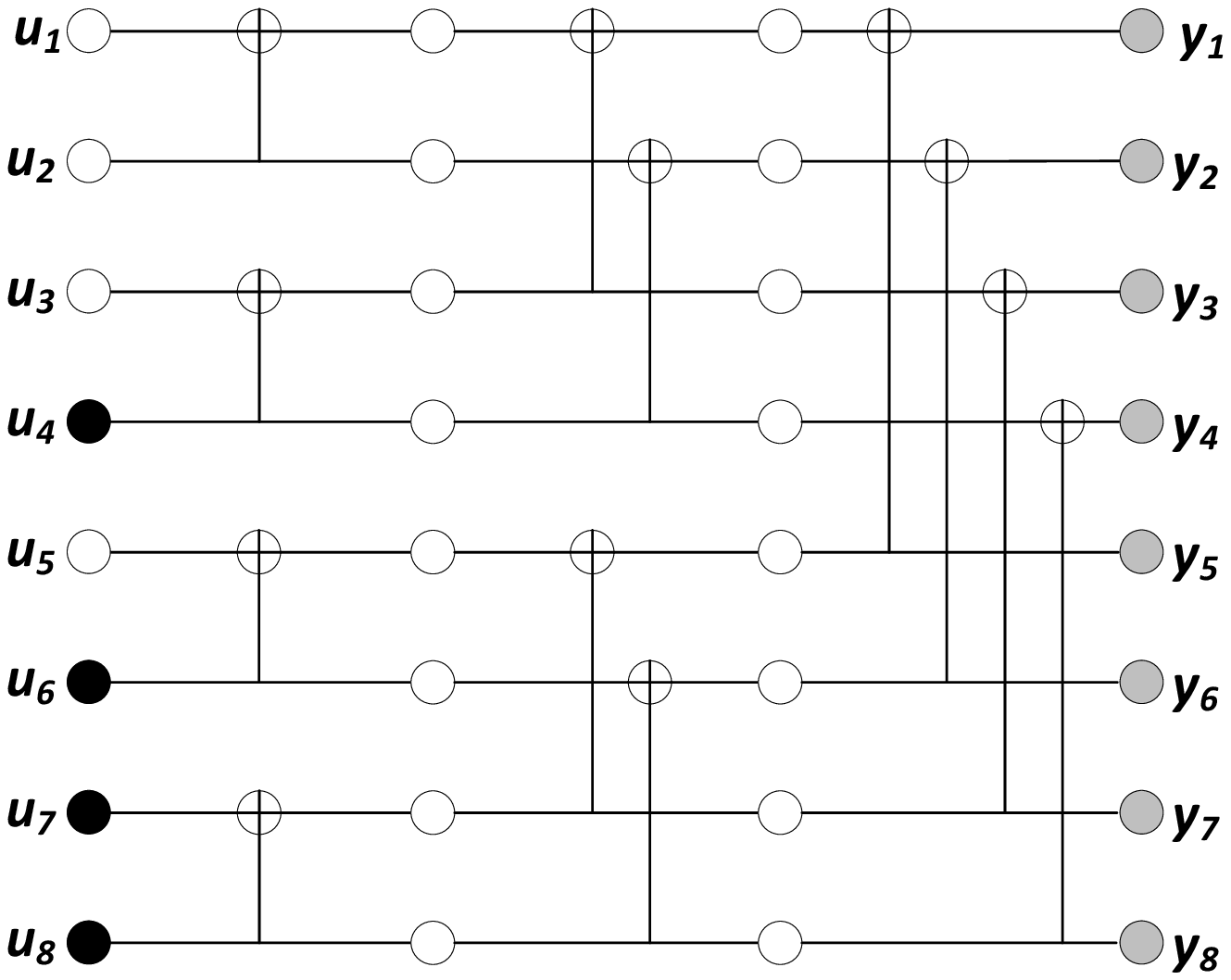}\label{encoder}}
\hfil
\subfloat[]{\includegraphics[width=0.8in]{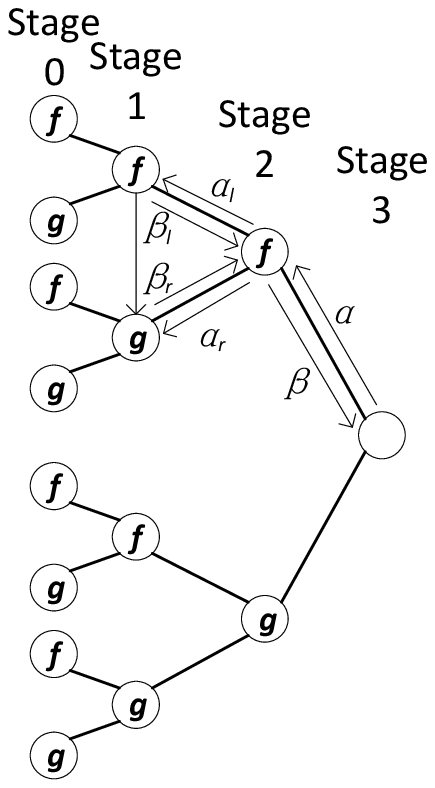}\label{SC_tree}}
\hfil
\caption{\protect\subref{encoder} Encoder of $(8,4)$ polar code, 
\protect\subref{SC_tree} Tree presentation of $(8,4)$ SC decoder
}
\label{polar_code_introduction}
\end{figure}

As introduced by E. Arikan~\cite{arikan2009channel}, we can construct polar code by successively performing channel polarization. Fig.~\ref{encoder} shows an example of the construction of 8-bit polar code.
Mathematically, polar codes are linear block codes of length $n = 2^m$. 
The coded codeword ${\bm{x}}\triangleq {(x_1,x_2,\cdots,x_n)}$ is computed by $\bm{x}=\bm{u}\bm{G}$ where $\bm{G=F^{\otimes m}}$, and $\bm{F^{\otimes m}}$ is the $m$-th Kronecker power of 
$\bm{F} = 
\begin{bmatrix}
1&0\\
1&1
\end{bmatrix}
$. 
Each row of $G$ is corresponding to an equivalent polarizing channel. 
For an $(n,k)$ polar code, $k$ bits that carry source information in $\bm{u}$ are called information bits. 
They are transmitted via the most $k$ reliable channels. 
While the rest $n-k$ bits, called frozen bits, are set to zeros and are placed at the least $n-k$ reliable channels. 

Polar codes can be decoded by recursively applying successive cancellation to estimate $\hat{u}_i$ from the channel output $y_{0}^{n-1}$ and the previously estimated bits $\hat{u}_{0}^{i-1}$. 
This method is referred as successive cancellation (SC) decoding.
Actually, SC decoding can be regarded as a binary tree traversal as described in Fig.~\ref{SC_tree}.
The number of bits of one node in stage $m (m = 0,1,2...log_2n)$ is equal to $2^m$. 
$\bm{\alpha}$ stands for the soft reliability value, typically is log-likelihood ratio (LLR). 
Each left and right child nodes can calculate the LLR for current node via $f$ and $g$ functions, respectively~\cite{leroux2013semi}. 
However, in order to solve $g$ function, a feedback $\bm{{\beta}_l}$ from left child of the same parent node is needed. This kind of feedback is called partial sum.
At stage 0, $\beta$ of a frozen node is always zero, and for information bit its value is calculated by threshold detection of the soft reliability according to 
 \begin{equation}\label{hard_decision}
  \beta=h(\alpha)=  
  \left
  \{
   \begin{array}{c}
   0,~if~\alpha \geqslant 0  \\
   1,~otherwise  \\
   \end{array}
  \right.
 \end{equation}  
At intermediate stages, $\bm{\beta}$ can be recursively calculated by
 \begin{equation}\label{feedback}
  \beta[i] =  
  \left
  \{
   \begin{array}{ll}
   \beta_{l}[i] \oplus \beta_{r}[i]~if~i\leq~N^{m}/2 \\
   \beta_{r}[i-N^{m}/2]~otherwise  \\
   \end{array}
  \right.
 \end{equation}

%Determining the location of the information and frozen bits depends on the channel model and the channel quality is investigated in~\cite{tal2013construct}.
%\begin{figure}[!tc]
%\centering
%\includegraphics[width=3in]{code_tree.eps}
%\caption{An example of LSC decoding with list size 4 for $(8,4)$ polar code from codeword tree aspect}
%\label{code_tree}
%\end{figure} 

\subsection{Constituent codes based SC decoding}
\label{Constituent codes based SC decoding}

SC decoding generally suffers from the high latency due to its inherent serial property.
The processing of obtaining the partial sum from each node significantly constrains the decoding speed.
Thus, in order to reduce the latency caused by partial sum calculation, constituent code based SC decoding has been proposed~\cite{alamdar2011simplified}, \cite{sarkis2014fast}. 
By finding some certain patterns in the source code, some part of the codeword and their corresponding partial sums can be estimated immediately without traversal.
This method significantly reduces the partial-sum-constrained latency. 
$\mathcal{N}^0$, $\mathcal{N}^1$, $\mathcal{N}^{SPC}$ and $\mathcal{N}^{REP}$ are the four commonest constituent code.

\begin{figure}[]
\centering
\subfloat[]{\includegraphics[width=1.2in]{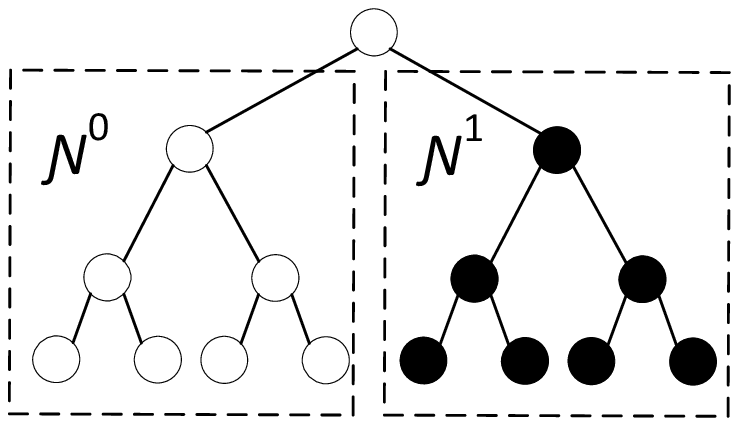}\label{n0n1}}
\hfil
\subfloat[]{\includegraphics[width=1.2in]{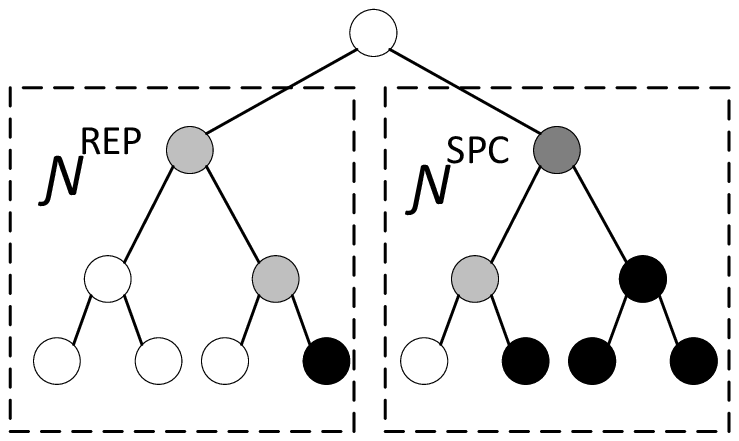}\label{nsnr}}
\caption{SC decoding tree simplified by constituent codes
%\protect\subref{n0n1} An example of $\mathcal{N}^0$ and $\mathcal{N}^1$ in a 8-bit %polar code tree, and \protect\subref{nsnr} An example of $\mathcal{N}^{SPC}$ and $%\mathcal{N}^{REP}$ in a 8-bit polar code tree
} 
\label{2tree}
\end{figure}

$\mathcal{N}^0$ and $\mathcal{N}^1$ only contain either frozen bits or information bits, respectively.
For $\mathcal{N}^0$ codes, we can set the corresponding partial sums to $0$ immediately. 
For $\mathcal{N}^1$ node, the partial sums can be directly determined via threshold
detection Eq.~(\ref{hard_decision}). 
$\mathcal{N}^{SPC}$ and $\mathcal{N}^{REP}$ contain both frozen bits and information bits.
In the $\mathcal{N}^{SPC}$ codes, only the first bit is frozen. It makes the length $n$ constituent codes as a rate $(n-1)/n$ single parity check (SPC) code.
This kind of code can be decoded by performing parity check with the least reliable bit. Typically it is the one with the minimum absolute value of LLR. 
In the $\mathcal{N}^{REP}$ codes, only the last bit is information bit.
In this case, all the corresponding partial sums should be the same since they all are the reflection of the last information bit. Thus, the decoding algorithm starts by summing all input LLRs and the partial sums are calculated by performing the hard detection to the final summary. Fig.~\ref{2tree} shows an example of how constituent code can simplify the SC decoding tree.  
According to T. Che's implementation of constituent code based SC decoder~\cite{che2016tc}, the latency of length $n$ constituent code can be reduced from $2n-2$ to 1, 1, $log_2n+1$ and $log_2n$ for $\mathcal{N}^0$, $\mathcal{N}^1$, $\mathcal{N}^{SPC}$ and $\mathcal{N}^{REP}$ codes, respectively.
In order to further optimize the performance constituent codes based decoder, a specific designed PSG for it is very necessary.   

%\subsection{Current existing partial sums generator}
%\label{Current existing partial sums generator}
%
%\subsubsection{Feed-forward partial sums generator}
%\label{Feed-forward partial sums generator}

\subsection{Shift-register-based partial sums generator}
\label{Shift-register-based partial sums generator}
\begin{figure}[!t]
\centering
\includegraphics[width=3in]{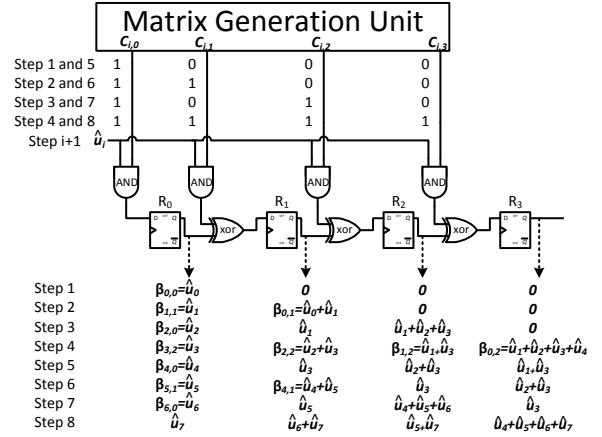}
\caption{The architecture of SR-PSG}
\label{SR_PSG}
\end{figure} 
 
Among all the aforementioned PSGs design, shift-register-based PSG (SR-PSG) has a better performance in terms of both the timing and hardware complexity. 
For length $n$ polar code decoder, it consists of $n$ registers and some other simple combination logic. Along with the estimation of each $\hat{u}_i$, the registers perform shift calculation and the partial sums can be obtained from their corresponding register. 
Its architecture is illustrated in Fig.~\ref{SR_PSG}. This architecture is built according to the following rule:
 \begin{equation}\label{SR_PSG_e}
  \left
  \{
   \begin{array}{l}
   R_0~\Leftarrow~\hat{u}_i\cdot c_{i,0} \\
   R_k~\Leftarrow~R_{k-1}\oplus (\hat{u}_i\cdot c_{i,k})  ,~if~k \geqslant 0 \\
   \end{array}
  \right.
 \end{equation}  
where $\cdot$ and $\oplus$ stand for $and$ and $exclusive$-$or$ operation, respectively.
In Fig.~\ref{SR_PSG}, $R_k$ means the $k$th register, $\hat{u}_i$ means the $i$th estimated bit. $\beta_{i,j}$ means the $j$th partial sum in stage $i$. $c_{i,k}$ means the $i$th row and $k$th column in the generate matrix $G$. The matrix generation unit is able to generate $c_{i,k}$ with very simple logic.   
The SC decoder consists of many basic computation parts called processing unit (PU). Each partial sum needs to be feed into the corresponding PU. The shift register based architecture can guarantee that all partial sum required by a PU are all generated in the same register, which can avoid any extra routing logic in the circuit. 

Such architecture is able to receive the estimated bit and update the corresponding partial sum by every valid cycle, which is highly consistent with SC decoding processing. However, this architecture is not suitable for constituent codes based SC decoder since some partial sums are obtained directly instead of calculating from estimated bits.
Thus, a PSG for constituent codes based SC decoder should have the capability to generate the new partial sums from either the directly got intermediate partial sums or the estimated bits, and to maintain the coherence of them.

\section{Proposed Design}
\label{Proposed Design}
%\begin{figure}[!tc]
%\centering
%\includegraphics[width=3in]{proposed_design.eps}
%\caption{The architecture of proposed design}
%\label{proposed_design}
%\end{figure} 
In this section, we first derive the mathematical presentation of constituent code based partial sum from Eq.~(\ref{SR_PSG_e}). Then, the overall hardware architecture and subsequent design details are presented.

\subsection{Mathematical Presentation}
\label{Mathematical Presentation}

\begin{figure}[!ht]
\centering
\subfloat[]{\includegraphics[width=1.5in]{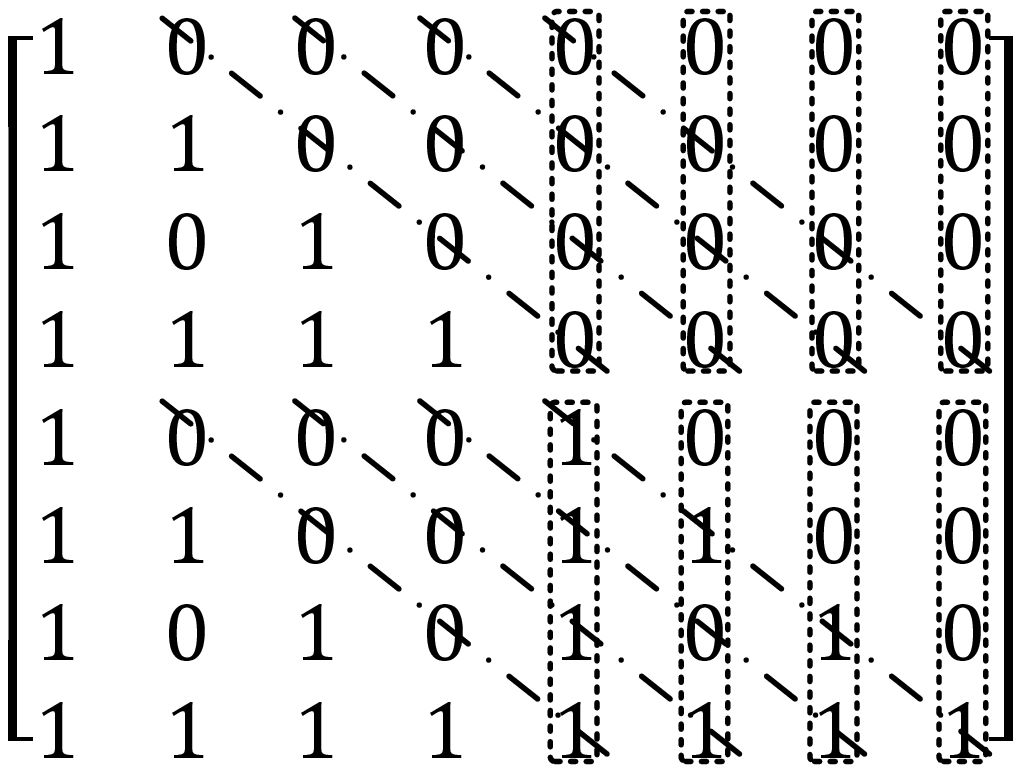}\label{observation}}
\hfil
\subfloat[]{\includegraphics[width=1.5in]{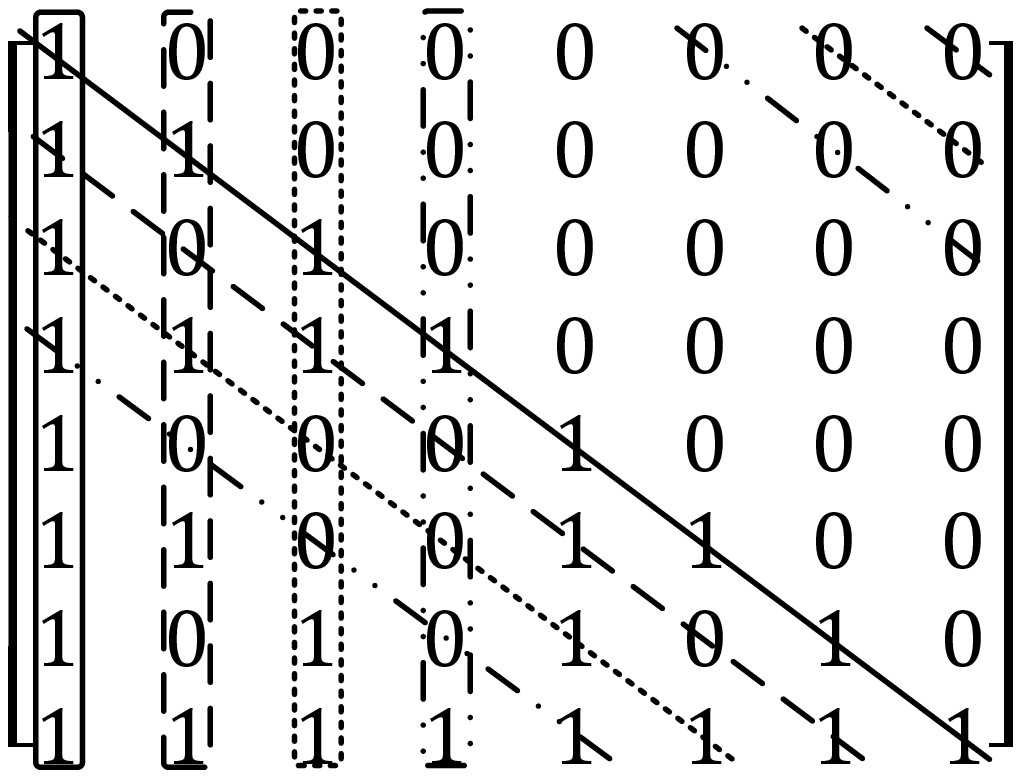}\label{diagonal_shift}}
\caption{
%SC decoding tree simplified by constituent codes
\protect\subref{observation} Elements shift in generation matrix, and \protect\subref{diagonal_shift} diagonal cycle-shift in generation matrix 
} 
\label{shift}
\end{figure}
For a length $n$ constituent code, its corresponding estimated bits and partial sums are denoted as $\hat{u}_{i-n+1}^c~\ldots~\hat{u_i}^c$ and $\beta_0^c\ldots \beta_{n-1}^c$, respectively.
All the $\bm{\beta^c}$ are obtained at the same time. 
For those bits that do not belong to any constituent codes, we still have to calculate their corresponding partial sums accroding to Eq.~(\ref{SR_PSG_e}). 
Thus, if we still want to keep consistency between directly calculated intermediate partial sums and the one-by-one-estimated bits, we need to derive the mathematical presentation with $\bm{\beta^c}$ from Eq.~(\ref{SR_PSG_e}).

For $k~\geqslant~n$ and $k\in[a\cdot n,(a+1)\cdot n-1], a~=~1,2,\ldots$, according to Eq.~(\ref{SR_PSG_e}), we have 
\begin{equation}
\begin{aligned}
\ R_k &= R_{k-1}\oplus (\hat{u}_i^c\cdot c_{i,k})\\ 
&= R_{k-2}\oplus (\hat{u}_{i-1}^c\cdot c_{i-1,k-1})\oplus (\hat{u}_i^c\cdot c_{i,k})\\
&\cdots\\
&=R_{k-n}\oplus\\
&\left(
\left[
\begin{array}{lll}
\hat{u}_{i-n+1}^c,&\cdots,&\hat{u}_i^c
\end{array}
\right]
\left[
\begin{array}{c}
c_{i-n+1,k-n+1}\\
\vdots\\
c_{i,k}
\end{array}
\right]
\right).
\label{klargerthann_1}
\end{aligned}  
\end{equation}
As we know, $c_{i,k}$ is the element of generation matrix $G$ which is the Kronecker power of 
$\bm{F} = 
\begin{bmatrix}
1&0\\
1&1
\end{bmatrix}
$. Combine this property with our observation on the matrix, we conduct the following rule which is also noted in Fig.~\ref{observation}.   
\begin{equation}
\left[
\begin{array}{c}
c_{i-n+1,k-n+1}\\
\vdots\\
c_{i,k}
\end{array}
\right]
=
\left[
\begin{array}{c}
c_{i-n+1,(a+1)\cdot n - (k~mod~n)-1}\\
\vdots\\
c_{i,(a+1)\cdot n - (k~mod~n)-1}
\end{array}
\right].
\label{shift_g}
\end{equation}

According to the definition of generation matrix and concept of constituent code, when $c_{i,k}~=~0$, the right part of Eq.~(\ref{shift_g}) is equal to a all $\bm{zero}$ vector, and when $c_{i,k}~=~1$ the right part of Eq.~(\ref{shift_g}) is equal to the $(n-(k~mod~n)-1)$th column in the generation matrix for length $n$ polar code. According to the definition of partial sum and Eq.~(\ref{feedback}), we get 
\begin{equation}
[\hat{u}_{i-n+1}^c,\cdots,\hat{u}_i^c]\cdot[c_{i-n+1,p(k)},\cdots,c_{i,p(k)}]^T=\beta_{p(k)}
\label{get_partial_2}
\end{equation}
where $p(k) = (n-(k~mod~n)-1)$.
%Thus, for $k\in[a\cdot n,(a+1)\cdot n-1], a~=~1,2,\ldots$, there is

Now we apply the above observation back to Eq~(\ref{klargerthann_1}). 
We define the vector $\bm{R_a}=[R_{a\cdot n},\cdots,R_{a\cdot n + n-1}]$ and $\bm{c_{i,a}} = [c_{i,a\cdot n},\cdots,c_{i,a\cdot n + n-1}]$ for $k\in[a\cdot n,(a+1)\cdot n-1], a~=~1,2,\ldots$.
We also define the vectors $\bm{\hat{u}^c}=[\hat{u}_{i-n+1}^c,\cdots,\hat{u}_i^c]$ and $\bm{\hat{\beta}^c}=[\beta_{n-1}^c,\ldots,\beta_0^c]$. Then, we have 
\begin{equation}
\begin{aligned}
\bm{R_a} &=[R_{a\cdot n},\cdots,R_{a\cdot n + n-1}] \\\\
&=[R_{(a-1)\cdot n},\cdots,R_{a\cdot n -1}]\oplus \\
&\left(
[\hat{u}_{i-n+1}^c,\cdots,\hat{u}_i^c]
\left[
\begin{array}{ccc}
c_{i-n+1,a\cdot n-n+1}	&\cdots	&c_{i-n+1,a\cdot n}	\\
\vdots					&\ddots	&\vdots				\\
c_{i,a\cdot n}			&\cdots	&c_{i,a\cdot n+n-1}
\end{array}
\right]
\right)\\
&=[R_{(a-1)\cdot n},\cdots,R_{a\cdot n -1}]\oplus \\
&\left(
\bm{\hat{u}^c}
\left[
\begin{array}{ccc}
c_{i-n+1,p(a\cdot n)}	&\cdots	&c_{i-n+1,p(a\cdot n+n-1)}	\\
\vdots					&\ddots	&\vdots				\\
c_{i,p(a\cdot n)}			&\cdots	&c_{i,p(a\cdot n+n-1)}
\end{array}
\right]
\right)\\
=&
  \left
  \{
   \begin{array}{ll}
   \bm{0},if~\bm{c_{i,a}}~=~\bm{0} \\
   \bm{R_{a-1}}\oplus\cdot\bm{\beta^c},if~\bm{c_{i,a}}~=~\bm{1}\\
   \end{array}
  \right.
\label{klargerthann_2}
\end{aligned}  
\end{equation}

For the consistent with Eq.~(\ref{SR_PSG_e}), we rewrite Eq.~(\ref{klargerthann_2}) as follow:
\begin{equation}
\bm{R_a}=\bm{R_{a-1}}\oplus(\bm{\beta^c} \& \bm{c_{i,a}})
\label{klargerthann_3}
\end{equation}
where $\&$ stands for the bit-wise $and$ operation.

For $0~\leqslant~k<~n$, similar to Eq.~(\ref{klargerthann_1}), we have
\begin{equation}
R_k =[\hat{u}_{i-k}^c,\cdots,\hat{u}_i^c]\cdot[c_{i-k,0},\cdots,c_{i,k}]^T
\label{ksmallerthann_1}
\end{equation} 
According to the definition of $G$ and constituent codes, we can conduct that for any length $n$ constituent codes, the first $n$ columns of its corresponding rows in $G$ should also be a generation matrix $G_n$ for length $n$ polar code. As described in Fig.~\ref{diagonal_shift}, the diagonal cycle shift is same as each correspond column, and consider the $G_n$ is a lower triangular matrix, we get
\begin{equation}
\begin{aligned}
&[c_{i-n+1,k+1},\cdots,c_{i-k-1,n-1},c_{i-k,0},\cdots,c_{i,k}]^T\\
&=[c_{i-n+1,n-1-k},\cdots,c_{i,n-1-k}]^T\\
&=[0,\cdots,0,c_{i-k,n-1-k},\cdots,c_{i,n-1-k}]^T
\end{aligned}
\label{cycle_shift}
\end{equation} 
Thus, Eq.~(\ref{ksmallerthann_1}) can be rewritten as:
\begin{equation}
\begin{aligned}
R_k 
&=[\hat{u}_{i-k+1}^c,\cdots,\hat{u}_i^c]\cdot[c_{i-k,0},\cdots,c_{i,k}]^T\\
&=[\hat{u}_{i-n+1}^c,\cdots,\hat{u}_i^c]\cdot[0,\cdots,0,c_{i-k,0},\cdots,c_{i,k}]^T\\
&=[\hat{u}_{i-n+1}^c,\cdots,\hat{u}_i^c]\cdot[c_{i-n+1,n-1-k},\cdots,c_{i,n-1-k}]^T\\
&=\beta_{n-k-1}^c
\label{ksmallerthann_2}
\end{aligned}
\end{equation} 
Thus, combining Eq.~(\ref{ksmallerthann_2}) and Eq.~(\ref{klargerthann_3}), we derive the mathematical presentation for partial sum of constituent based polar decoder as follow:
 \begin{equation}\label{psg_cbpc}
  \bm{R_a} =  
  \left
  \{
   \begin{array}{ll}
   \bm{\beta^c},~if~a~=~0 \\
   \bm{R_{a-1}}\oplus(\bm{\beta^c} \& \bm{c_{i,a}}), if~a\geqslant~1.  \\
   \end{array}
  \right.
 \end{equation}

\subsection{Proposed architecture}
\label{Proposed architecture}

\begin{figure}[!t]
\centering
\includegraphics[width=3.5in]{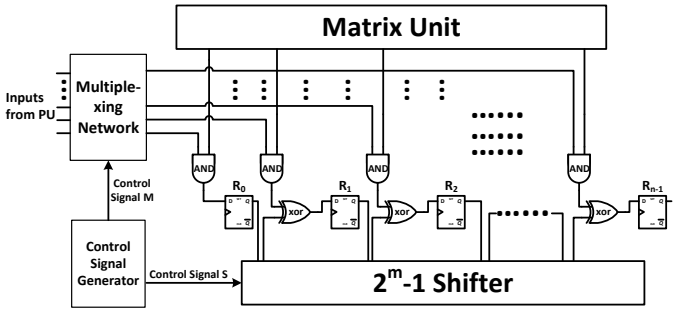}
\caption{Overall architecture of SR-CB-PSG}
\label{SR_CB_PSG}
\end{figure}

According to Eq~(\ref{psg_cbpc}), the shift-register constituent-code based partial sum generator (SR-CB-PSG) is proposed as in Fig.~\ref{SR_CB_PSG}.
Compared with Fig.~\ref{SR_PSG}, there are three differences. 
The first difference is the input. 
For SR-PSG, only current estimated bit is sent into, which means the input is only from the $PU$ from stage $0$.
However, for SR-CB-PSG, the inputs are from $PU$s of any stage, depending on the length of constituent code.
Thus, a multiplexing networking is needed to route all the inputs values to the right registers.
The second difference is the shift function. 
According to Eq~(\ref{psg_cbpc}), instead of just shifting by one bit, the shifter should have the capability to shift $n$-bit where $n$ is the length of constituent code. According to the definition of constituent code, $n$ should be the any power of $2$.
Thus, A specific design $(2^m-1)$-bit shifter is proposed. 
The control signals for both the muxing networking and shifter are from the $Control~Signal~Generator (CSG)$ with simple logic.
The last difference is matrix generation unit. For each constituent code, its corresponding $c_{i,j}$ should be the $i$th row of the generation matrix, where $i$ is the index of the last bit in the constituent code. Due to the irregularity of the constituent code, it's unnecessary to build an online generator for that. 
Thus, a pre-calculated ROM is placed. It is a trade-off between design complexity and hardware resource. It can be replaced by a re-configurable memory device like RAM for flexibility.

\begin{figure}[!t]
\centering
\includegraphics[width=3.5in]{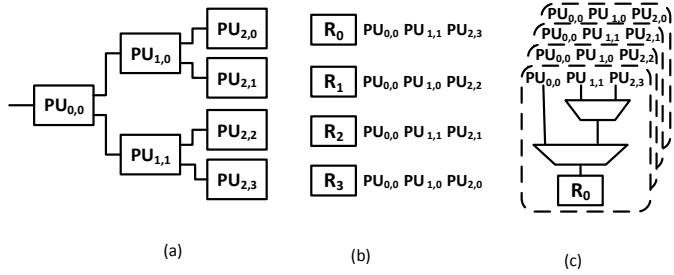}
\caption{
(a) PU tree of SC decoder, (b) PUs and their corresponding register, and (c) architecture of multiplexing network    
}
\label{mux_networking}
\end{figure}

%\begin{figure}[!ht]
%\centering
%\subfloat[]{\includegraphics[width=1in]{mn1.eps}\label{mn1}}
%\hfil
%\subfloat[]{\includegraphics[width=0.7in]{mn2.eps}\label{mn2}}
%\hfil
%\subfloat[]{\includegraphics[width=0.5in]{mn3.eps}\label{mn3}}
%\caption{
%%SC decoding tree simplified by constituent codes
%\protect\subref{mn1} PU tree of SC decoder, \protect\subref{mn2} PUs and their corresponding register, and \protect\subref{mn3}architecture of multiplexing network    
%}
%\label{mux_networking}
%\end{figure}

Fig.~\ref{mux_networking} shows an example of partial sum routing for 8 bit constituent code based polar code. 
We can see each register has specific corresponding PU from each stage. 
They need the multiplexing networking to route the partial sums to the each right register. 
For length $n$ polar code, there are $log_2n$ stages in the decoder and $n/2$ registers in the SR-CB-PSG.  
If the multiplexing networking is built from the basic 2-bit MUX, each register is assigned an identical MUXs networking made by $(log_2n~-~1)$ MUXs.
All the networkings share the same control signal. According to its architecture, the control signals are the direct binary mapping of its stage index. 
In total, $n/2\cdot(log_2n~-~1) $ MUXs are needed. 
Since the multiplexer networking needs to wait each PU finish computing to get the valid inputs, it is on the critical path of the decoder.
Thus, it causes additional $\lceil log_2(log_2n)\rceil \cdot \bigtriangleup(MUX)$ delay, where $\bigtriangleup(MUX)$ is the delay for a single MUX. 
      
\begin{figure}[!t]
\centering
\includegraphics[width=3in]{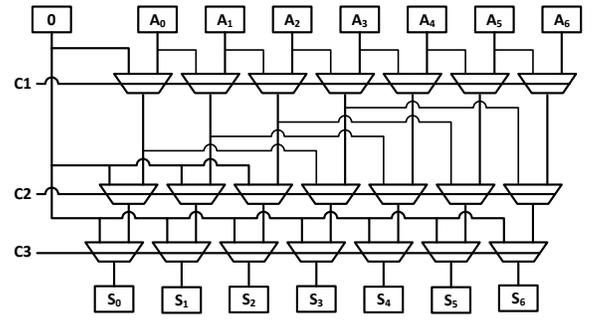}
\caption{An example of $(2^m-1)$ shifter for 16-bit polar code decoder}
\label{shifter}
\end{figure}

For the $(2^m-1)$ shifter, we proposed a barrel-shifter-based architecture. For length $n$ polar code, $m\leqslant (log_2n-1)$.  
The shifter performs logic right shift. 
For $k<n$, where $k$ is the index of the register and $n$ is the length of the current constituent code, zeros are added to the left. 
For $k\geqslant n$,  we do shift. Those behaviors satisfy the first and second in Eq~(\ref{psg_cbpc}).  

Fig.~\ref{shifter} shows an example of $(2^m-1)$ shifter for 16-bit polar code decoder. All the MUXs in the same row can shall the same control signal.
Those signals are generated by a $k~to~2^k~decoder$, where $k=\lceil log_2(log_2n) \rceil$ for length $n$ polar code.
For length $n$ polar code, there are $(n/2-1)\cdot(log_2n~-~1) $ MUXs are needed for the shifter.
Since the shifter can start shift data without waiting $PU$ to finish computing, it is not on the critical path.
Thus, it should not deteriorate the timing performance of the decoder at all.

\section{Implementation results and discussions}
\label{Implementation results and discussions}

To the best of our knowledge, the proposed design is the first PSG design especially design for constituent codes based SC decoder.
Thus, there is no reference design we can directly compare with. 
In this section, we list all the results we have and presents some relevant discussions.

%\begin{table}[]
%\centering
%\caption{Critical Path Comparison}
%\label{Critical Path Comparison}
%\begin{tabular}{lll}
%               	& Proposed 	& \cite{berhault2013partial} \\
%Critical Path 	& $\lceil log(logN)\rceil \cdot \bigtriangleup(MUX) +  \bigtriangleup(AND) + \bigtriangleup(XOR)$		& $ \bigtriangleup(AND) + \bigtriangleup(XOR)	$
%\end{tabular}
%\end{table}

\begin{table}[h]
\centering
\caption{Critical Path Comparison}
\label{Critical Path Comparison}
\begin{tabular}{@{}c|c@{}}
\hline
                             	& Critical Path                                                                                  \\ \hline
SR-PSG\cite{berhault2013partial} 		& $ \bigtriangleup(AND) + \bigtriangleup(XOR)	$                                                  \\  	
Proposed                     	& $\lceil log(logN)\rceil \cdot \bigtriangleup(MUX) +\bigtriangleup(AND) + \bigtriangleup(XOR)$   \\ \hline                            

\end{tabular}
\end{table}
Table~\ref{Critical Path Comparison} shows the critical path comparison between proposed PSG and the PSG in~\cite{berhault2013partial}. We can tell the delay overhead comes from the muxing network. Ideally, the maximum frequency of constituent codes based SC decoder is lower than that of conventional SC decoder. 
However, after taking the latency reduction into account, as shown in Table~\ref{latency_reduction}, constituent codes based SC decoder is able to achieve much higher throughput.
The conventional SC decoder is referred from~\cite{yuan2014low} which is the lowest latency conventional SC decoder to the best of out knowledge.

\begin{table}[]
\centering
\caption{Decoder Latency comparison for length=1024 polar code}
\label{latency_reduction}
\begin{tabular}{c|ccccc}
\hline
                                        & \multicolumn{5}{c}{code rate}   \\ \cline{2-6} 
                                        & 0.2  & 0.35 & 0.5  & 0.65 & 0.8  \\ \hline
latency of conventional \cite{yuan2014low} & \multicolumn{5}{c}{767}         \\ 
latency of constituent code based                  & 263  & 298  & 266  & 200  & 160  \\ \hline

reduction($\%$)                               & 65.7 & 61.1 & 65.3 & 73.9 & 79.1 \\ \hline
\end{tabular}
\end{table}

Table \ref{Estimated resource consumption } shows the resource consumption estimation of proposed SR-CB-PSG for length $n$ polar code decoder and the comparison with other two conventional PSG. The most resource consumption part is the $MUX$ since it used in both multiplexer networking and shifter. 
The estimation for the ROM size is based on the average calculation since the decoding latency changes along with the code rate.

\begin{table}[h]
\centering
\caption{Resource comparison }
\label{Estimated resource consumption }
\begin{tabular}{@{}c|ccc@{}}
\toprule
%							&DFF			& MUX           		& XOR   	& AND 	& ROM(bit)      	\\ \midrule
%proposed 					&$N/2$ 			& $(N-1)\cdot(logN-1)$ 	& $N/2-1$ 	& $N/2$ & $N^2/10$(average) \\ 
%\cite{berhault2013partial} &$N$ 			& -						& $N-2$ 	& $N/2$ & 					\\ 
%\cite{zhang2013low} 		&$(N^2-4)/12$ 	& $N-2$					& $N/2-1$ 	&	-	& 	-				\\ \bottomrule
		&proposed				& \cite{berhault2013partial}    & \cite{zhang2013low} 	\\ \hline
DFF		&$n/2$ 					&$n$ 							&$(n^2-4)/12$ 			\\
MUX		&$(n-1)\cdot(logn-1)$ 	&-								&$n-2$					\\
XOR		&$n/2-1$				&$n-2$							&$n/2-1$				\\
AND		&$n/2$					&$n/2$							&-						\\
ROM(bit)&$n^2/10$(average)		&-								&-						\\ \bottomrule
\end{tabular}
\end{table}

The proposed design can be targeted on either ASIC or FPGA. 
We synthesized both with Nangate FreePDK 45nm process and on Xilinx Kintex-7 FPGA KC705 Evaluation board. 
Table~\ref{hardware resource of SR-CB-PSG} shows the hardware resource of SR-CB-PSG for 1024 code length polar code decoder on both of them. 

%The additional delays are also listed in table~\ref{hardware resource of SR-CB-PSG}. 
%As discussed in Sec.~\ref{Proposed architecture}, the delay overhead is from the multiplexer networking. 
%Compared with the  critical path list in \cite{che2016tc}, the delay overhead is obviously negligible. 
\begin{table}[h]
\centering
\caption{hardware resource of SR-CB-PSG for 1024 code length polar code decoder }
\label{hardware resource of SR-CB-PSG}
\begin{tabular}{c|cc|c}
\toprule
                                   & \multicolumn{2}{c|}{XC7K325T-2FFG900C FPGA} & nangate 45nm \\ \hline
\multirow{2}{*}{Hardware Resource} & slice LUTs            & slice REGs          & area         \\
                                   & $1569(\textless1\%)$    & $512(\textless1\%)$   & $16333\mu m^2$        \\ \bottomrule
                                   %\hline
%Additional Delay                
\end{tabular}
\end{table}    

Noticeably, the architecture we discussed in this paper is based on the consideration for the worst case, which is that the maximum length of constituent codes could be $n/2$. 
However, for practical application, the maximum length of constituent is fix for certain code rate and usually cannot approach $n/2$. 
For those case, the logic of both the multiplexer networking and shifter could be even simpler, which will result in a better timing and silicon area performance.

\section{Conclusion}
\label{Conclusion}

This paper proposed an efficient PSG hardware design for constituent code based SC decoder. 
Conventional PSG is not compatible with the constituent code based SC decoder.
This is because that the conventional one is only capable of taking estimated bit one by one but the constituent code based decoder is generated the intermediate partial sum directly.
To solve this problem, we first derive the mathematical presentation for constituent code based PSG from the SR-PSG for conventional SC decoder. 
Then, the overall hardware architecture and design details are proposed.
Finally, the implementation result with both VLSI and FPGA technology are presented, and the relevant discussions are carried out.
%and is able to achieve a good timing performance. 

\bibliographystyle{IEEEtran}
\bibliography{IEEEabrv}

\begin{thebibliography}{10}
\providecommand{\url}[1]{#1}
\csname url@samestyle\endcsname
\providecommand{\newblock}{\relax}
\providecommand{\bibinfo}[2]{#2}
\providecommand{\BIBentrySTDinterwordspacing}{\spaceskip=0pt\relax}
\providecommand{\BIBentryALTinterwordstretchfactor}{4}
\providecommand{\BIBentryALTinterwordspacing}{\spaceskip=\fontdimen2\font plus
\BIBentryALTinterwordstretchfactor\fontdimen3\font minus
  \fontdimen4\font\relax}
\providecommand{\BIBforeignlanguage}[2]{{%
\expandafter\ifx\csname l@#1\endcsname\relax
\typeout{** WARNING: IEEEtran.bst: No hyphenation pattern has been}%
\typeout{** loaded for the language `#1'. Using the pattern for}%
\typeout{** the default language instead.}%
\else
\language=\csname l@#1\endcsname
\fi
#2}}
\providecommand{\BIBdecl}{\relax}
\BIBdecl

\bibitem{arikan2009channel}
E.~Arikan, ``Channel polarization: A method for constructing capacity-achieving
  codes for symmetric binary-input memoryless channels,'' \emph{Information
  Theory, IEEE Transactions on}, vol.~55, no.~7, pp. 3051--3073, 2009.

\bibitem{tal2011list}
I.~Tal and A.~Vardy, ``List decoding of polar codes,'' in \emph{Information
  Theory Proceedings (ISIT), 2011 IEEE International Symposium on}.\hskip 1em
  plus 0.5em minus 0.4em\relax IEEE, 2011, pp. 1--5.

\bibitem{hussami2009performance}
N.~Hussami, S.~B. Korada, and R.~Urbanke, ``Performance of polar codes for
  channel and source coding,'' in \emph{Information Theory, 2009. ISIT 2009.
  IEEE International Symposium on}.\hskip 1em plus 0.5em minus 0.4em\relax
  IEEE, 2009, pp. 1488--1492.

\bibitem{xu2015xj}
J.~Xu, T.~Che, and G.~Choi, ``Xj-bp: Express journey belief propagation
  decoding for polar codes,'' in \emph{2015 IEEE Global Communications
  Conference (GLOBECOM)}.\hskip 1em plus 0.5em minus 0.4em\relax IEEE, 2015,
  pp. 1--6.

\bibitem{yuan2013architecture}
B.~Yuan and K.~K. Parhi, ``Architecture optimizations for bp polar decoders,''
  in \emph{Acoustics, Speech and Signal Processing (ICASSP), 2013 IEEE
  International Conference on}.\hskip 1em plus 0.5em minus 0.4em\relax IEEE,
  2013, pp. 2654--2658.

\bibitem{lin2016high}
J.~Lin, J.~Sha, L.~Li, C.~Xiong, Z.~Yan, and Z.~Wang, ``A high throughput
  belief propagation decoder architecture for polar codes,'' in \emph{Circuits
  and Systems (ISCAS), 2016 IEEE International Symposium on}.\hskip 1em plus
  0.5em minus 0.4em\relax IEEE, 2016, pp. 153--156.

\bibitem{leroux2013semi}
C.~Leroux, A.~J. Raymond, G.~Sarkis, and W.~J. Gross, ``A semi-parallel
  successive-cancellation decoder for polar codes,'' \emph{Signal Processing,
  IEEE Transactions on}, vol.~61, no.~2, pp. 289--299, 2013.

\bibitem{zhang2013low}
C.~Zhang and K.~Parhi, ``Low-latency sequential and overlapped architectures
  for successive cancellation polar decoder,'' \emph{Signal Processing, IEEE
  Transactions on}, vol.~61, no.~10, pp. 2429--2441, 2013.

\bibitem{yuan2014low}
B.~Yuan and K.~K. Parhi, ``Low-latency successive-cancellation polar decoder
  architectures using 2-bit decoding,'' \emph{IEEE Transactions on Circuits and
  Systems I: Regular Papers}, vol.~61, no.~4, pp. 1241--1254, April 2014.

\bibitem{che2015overlapped}
T.~Che, J.~Xu, and G.~Choi, ``Overlapped list successive cancellation approach
  for hardware efficient polar code decoder,'' \emph{arXiv preprint
  arXiv:1511.00577}, 2015.

\bibitem{balatsoukas2014hardware}
A.~Balatsoukas-Stimming, A.~J. Raymond, W.~J. Gross, and A.~Burg, ``Hardware
  architecture for list successive cancellation decoding of polar codes,''
  \emph{Circuits and Systems II: Express Briefs, IEEE Transactions on},
  vol.~61, no.~8, pp. 609--613, 2014.

\bibitem{balatsoukas2014llr}
A.~Balatsoukas-Stimming, M.~Bastani~Parizi, and A.~Burg, ``Llr-based successive
  cancellation list decoding of polar codes,'' in \emph{Acoustics, Speech and
  Signal Processing (ICASSP), 2014 IEEE International Conference on}.\hskip 1em
  plus 0.5em minus 0.4em\relax Ieee, 2014, pp. 3903--3907.

\bibitem{lin2015hybrid}
J.~Lin and Z.~Yan, ``A hybrid partial sum computation unit architecture for
  list decoders of polar codes,'' in \emph{2015 IEEE International Conference
  on Acoustics, Speech and Signal Processing (ICASSP)}.\hskip 1em plus 0.5em
  minus 0.4em\relax IEEE, 2015, pp. 1076--1080.

\bibitem{berhault2013partial}
G.~Berhault, C.~Leroux, C.~Jego, and D.~Dallet, ``Partial sums generation
  architecture for successive cancellation decoding of polar codes,'' in
  \emph{SiPS 2013 Proceedings}.\hskip 1em plus 0.5em minus 0.4em\relax IEEE,
  2013, pp. 407--412.

\bibitem{berhault2015partialsum}
------, ``Partial sums computation in polar codes decoding,'' in \emph{2015
  IEEE International Symposium on Circuits and Systems (ISCAS)}.\hskip 1em plus
  0.5em minus 0.4em\relax IEEE, 2015, pp. 826--829.

\bibitem{fan2014efficient}
Y.~Fan and C.-y. Tsui, ``An efficient partial-sum network architecture for
  semi-parallel polar codes decoder implementation,'' \emph{IEEE Transactions
  on Signal Processing}, vol.~62, no.~12, pp. 3165--3179, 2014.

\bibitem{alamdar2011simplified}
A.~Alamdar-Yazdi and F.~R. Kschischang, ``A simplified successive-cancellation
  decoder for polar codes,'' \emph{IEEE communications letters}, vol.~15,
  no.~12, pp. 1378--1380, 2011.

\bibitem{sarkis2014fast}
G.~Sarkis, P.~Giard, A.~Vardy, C.~Thibeault, and W.~J. Gross, ``Fast polar
  decoders: Algorithm and implementation,'' \emph{Selected Areas in
  Communications, IEEE Journal on}, vol.~32, no.~5, pp. 946--957, 2014.

\bibitem{che2016tc}
T.~Che, J.~Xu, and G.~Choi, ``Tc: Throughput centric successive cancellation
  decoder hardware implementation for polar codes,'' in \emph{2016 IEEE
  International Conference on Acoustics, Speech and Signal Processing
  (ICASSP)}.\hskip 1em plus 0.5em minus 0.4em\relax IEEE, 2016, pp. 991--995.

\end{thebibliography}

\end{document}